# On Female Audience Sending Virtual Gifts to Male Streamers on Douyin

Huilian Sophie Qiu

## Introduction

Live streaming has become popular in the past few years. While many platforms overseas allow users to pay to subscribe to streamers, e.g., Twitch, Chinese live-streaming platforms, such as YY, Huya, and Douyin, provide a new way for streamers to earn money by receiving virtual gifts from the audience. This is a mechanism different from the subscription system. Sending virtual gifts is a function during live-stream sessions. The audience can send an arbitrary value of virtual gifts multiple times throughout a live-stream session. By now, the only way for content creators to directly earn money through Chinese live-streaming platforms is via the mechanism of the virtual gift (Kaye 2020). This mechanism makes livestreamer a new and popular occupation.

As a nascent mechanism, sending virtual gifts has become an interesting phenomenon to study (Zhang 2019). While there are several exploratory studies on why people send virtual gifts, they treat all kinds of live streams as a whole. We argue that such a coarse-grain approach misses out on the diversity of live stream contents. For example, there are live streams for small businesses, chatting, singing, or games. There are live streams done by streamers of different gender or age groups. These different types of live streams aim at different audience groups, who have different psychology and financial backgrounds. The content of the live stream also affects how people send virtual gifts. For example, it is unlikely that people will send tons of virtual gifts to live streams done by small businesses because these streamers earn money by selling products.

Our study is interested in why female audiences send virtual gifts to male streamers. There are several reasons why this topic needs attention. Firstly, while the male gaze has been around for a long while, the female gaze is on the rise. There are more female commodities that use young male idols in their advertisements (Li 2020). Female audiences watching a young man's live stream is an example of this reversed gender relationship. Moreover, the presence of a large number of female audiences who are willing to spend a lot of money on a young man reflects Chinese women's increasing financial power. Furthermore, while these female audiences may enjoy themselves when watching live streams, these virtual events may not have challenged the patriarchal structure. Are these women using these online platforms as a utopia?

In this paper, we aim to answer these three research questions:

RQ1: Why do female audiences watch these live streams by young male streamers on Douyin?
RQ2: Why do female audiences send virtual gifts to male streamers?
RQ3: What is the relationship between female audiences and male streamers?

We use Douyin, one of the most popular short video platforms, as a case study. By April 2020, Douyin and TikTok (Douyin's international sibling) have more than 2 billion downloads (Qu 2020). Although live streaming is not Douyin's primary functionality, it provides us the potential to study the relationship between short videos and live streams, which we would like to pursue as future work.

# Background

## Live streams on Douyin

Started as an app for short videos, Douyin also allows users to live stream. On the 'for you' page, users may see recommended short videos or live streams. On the upper left corner, there is an entry to the live stream plaza, where users can find live streams by category, such as e-commerce, games, music, or chatting.

In live stream rooms, the audience may choose to send virtual gifts to the streamer. One Chinese Yuan (~$0.16) is 10 Douyin coins. The cheapest gift is worth 0.1 Chinese Yuan (~$0.02) and the more expensive ones can exceed 3000 Chinese Yuan (~$466.05). More expensive gifts usually have fancier animations. Douyin takes 50% of a streamer's virtual gift income.

Many streamers have contracts with streamer guilds that can help them make short videos, advertise their live streams (by acquiring slots on Douyin's live stream plaza), and train them to become a streamer. In return, these guilds take 5% of the streamers' income and the streamer takes the rest 45%. These guilds usually have requirements for minimum hours and minimum virtual gifts. For example, one streamer told us that he has to receive 60,000 Chinese Yuan (~$9321) per month to get his company to provide him more resources, such as, recommendation on live stream plaza. Zhang et al. has a more comprehensive introduction on these streamer guilds (Zhang et al. 2019).

Each streamer has a fan group on Douyin. The audience can spend 0.1 Chinese Yuan to join the fan group. Each day, fans can increase their fan level by performing some tasks: spend 0.1 Chinese Yuan to send a fan's board, watch the live stream for a certain amount of time, and send a certain amount of virtual gifts. Many streamers create a fan group on Wechat or Weibo. The requirements for entering these external fan groups can be having reached a certain fan level, having sent a certain amount of virtual gifts within a day, or having joined the Douyin fan group for a certain amount of time. External fan groups are a way for streamers and their fans to interact outside of the live stream room.

## Virtual gifts on Douyin

While watching live streams is free, Douyin has devised various ways to incite the audience to send virtual gifts. When one enters a live stream room, after watching a while, Douyin will show a banner at the bottom of the screen, saying, "please send me a heart," which is one of the cheapest gifts (0.1 Chinese yuan), as shown in the figure below. Every now and then, Douyin adds new virtual gifts. Some are for special occasions, such as the spring equinox or some festival. Some are for special events, where users can unlock new gifts by sending enough certain gifts.

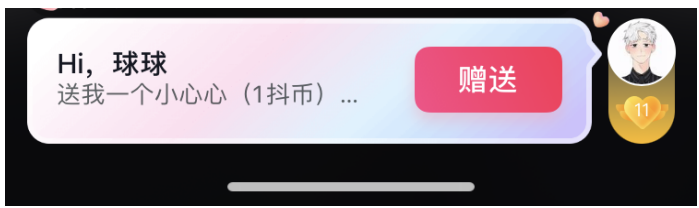

In addition, streamers have ways to disclose their wishes. Each streamer can choose gifts he wants to receive that day. A couple of months ago, Douyin added a gift exhibition hall, where people can see what gifts this streamer has received this week and which user has sent how many. Streamers can also write their wish list on the screen, as shown in the figure below.

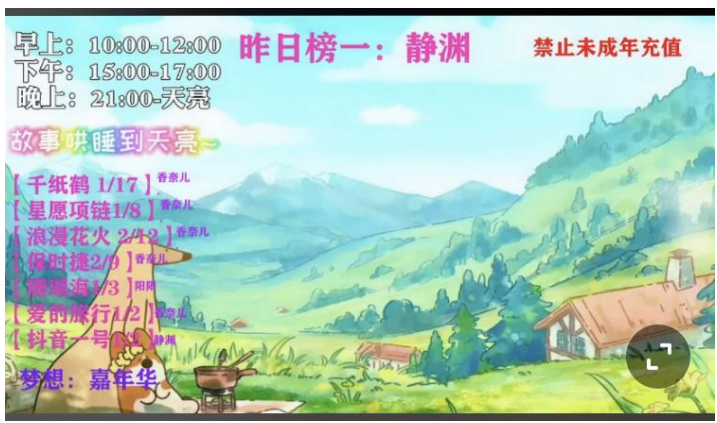

Douyin also allows two streamers PK with each other as a way to encourage the audience to spend more money. Throughout the PK, usually, 5 minutes long, the streamer with fewer gifts has to do certain things, such as shouting "help" or staying silent. After the PK, the streamer with fewer gifts has to do certain punishments, such as putting up a funny effect on his face or doing 30 push-ups. Therefore, sending virtual gifts is a way to make the other streamer do something.

Virtual gifts streamers receive during a PK are the same as those received otherwise; streamers can still earn 45%-50% of their value. Therefore, even though a streamer loses a competition, he has still earned money. Sometimes, the audience may get very upset and the streamer would comfort them.

## Gamification

Douyin has several gamification features in its UI designs.

It is very easy to see who has spent how much money and who spent the most. Users all have badges in front of their names reflecting the amount of money they have spent (wealth level) and their fanship (fan group name and fan level). In addition, there is a leaderboard (shown below) in every live stream room recording the amount of money each present user spent during that session.

Spending money also allows one to gain more attention. People with higher wealth levels can enjoy fancy animations when they enter a live stream room. Badges of people with lower fan levels are more transparent and of lighter colors than those of higher levels. Only the top 3 on the leaderboard can have their profile pictures displayed on the interface.

| | 在线观众 | |
|---|---|---|
| 观众信息 | | 音浪 |
| 1 | 洛笙 🔥 27 | 1.0万 |
| 2 | 碰碰的乖乖 51 2 洛府 | 6401 |
| 3 | 小球藻 22 10 洛府 | 1328 |
| 4 | 一笙独一☀夏夏… 40 18 洛府 | 766 |
| 5 | 一笙一世✨诺 38 17 洛府 | 101 |

Leaderboard and users' badges

# Related work

The behavior of watching live streams and sending virtual gifts has attracted scholars' attention in recent years. Using surveys and interviews, Lu et al. (2018) described several aspects of live streaming practices, such as contents, interactions, such as virtual gifts and fan groups, and elements of engaging live streams. Li et al. (Li 2018) attempted to use the flow theory[1] to analyze the psychological factors of sending virtual gifts and found that interactions and social presence are associated with the intention of sending virtual gifts. They also found that, when the flow is high, men have higher consumption intentions than women. Zhu et al. (2017) used trace data to unearth several behavior patterns of the audience sending virtual gifts. For example, a longer viewing time is associated with a higher intention of sending virtual gifts. Also, seeing others sending virtual gifts may stimulate viewers to send gifts. They also found that

Zhang et al. (2019) described the relationships between streamers, especially the ones with more followers, and their audience as "commodified virtual relationships." From their interviews and observations, on one hand, Zhang et al. found that as streamers gain more followers, they feel less emotionally attached to them, rather, they may consider the audience as possesseble, a source of revenue, and their relationships are "not an experience of affinity and closeness, but work that is frustratingly soulless and impersonal." On the other hand, some audiences consider sending virtual gifts as a way to enhance their social status among the fans and some even may treat streamers, especially female streamers, as "objects of the gaze, desire, and consumption."

However, these studies are looking at all types of live streams and audiences. Our study is interested in female audiences who watch young male streamers. Zhang et al. (2019) reported that many female streamers are being objectified by the audience, which is unsurprising given the prevalence of the male gaze. The relationship between female viewers and male content providers is the opposite of the male gaze. What are female audiences seeking from young male streamers? Lu and Lu (2019) reported that one of the reasons why young females engage with short videos is to have "a virtual 'intimate' relationship with their favorite content providers." Li (2020) analyzed the female gaze in mainland China and its implication on female consumers' self-empowerment. Do female audiences feel empowered by sending virtual gifts?

Currently, we have focused on streamers whose live stream rooms have fewer than 100 audiences, as they are more likely to develop closer relationships with the audience.

1. Flow is defined to be "the psychological state when people are experiencing intrinsic enjoyment with total involvement in the activity at hand" (Li et al. 2018)

## Methods

### Data collection

To answer our research questions, we conducted semi-interviews with some seasoned audiences. For each RQ, we design 3 main questions and several follow-up questions. The interviews take around 25 minutes to complete.

We recruited participants from Wechat fan groups and by sending direct messages to audiences with high rankings on the leaderboard. The first strategy proves to be the most effective. These Wechat fan groups are for fans with 10 fan levels or higher and the sizes are around 20 people. To achieve such high fan levels, one needs to visit the live stream room for more than a month, so these are very dedicated fans. Since the author is also a member of the groups, these interviewees have known her well and therefore are more likely to accept our interview invitation. We recruited 10 participants from 3 streamers' Wechat groups. Two of the streamers have fewer than 10K followers and one has more than 3M.

However, on Douyin, messages from strangers are all merged into one entry. As a result, it is hard for strangers to see my message and many people have shut off the messaging function or never read any messages from strangers. So far, we have only successfully recruited 2 participants from Douyin strangers. We also found an interviewee introduced by a streamer.

In the end, we conducted 13 interviews. Only two of them were conducted via audio chats, because most people are either shy or consider audio chat too much work. Interviews were all conducted in Chinese. Interviewers were not compensated.

### Data analysis

One author coded 6 interviews using an open coding method (Corbin et al. 2014). Before coding, the author used Google translation to translate interview transcripts into English. The coding was performed in English. Codes were later merged into themes. We will only report findings from the 6 interviews in this report. All these 6 participants are dedicated fans to a streamer who has a nice voice but does not show his face.

# Results

## Participants

We first report the demographics of the participants. All 6 participants are living in second- or third-tier cities in China. Among the ones who disclosed their ages, 2 are above 30 years of age and 3 are between 20 and 30. Four of them are single, one has a boyfriend, and one did not disclose.

These are all very dedicated fans who have spent a tremendous amount of time and money on the streamers. They all have 10 or higher fan levels. Most of them started watching live streams in 2019 or 2020. One started in January 2021. They all have wealth levels higher than 30, which means they have spent at least 6,000 Chinese Yuan (~$932). All of them say they would stay the entire live stream session, which can be 3 hours or longer per day. P14 confessed that, if she is busy, she will "at least try to finish the task (watch 20 minutes of live stream)" (P14). Some say they would do other things during the live streaming (P14, P11), some say they would "try [their] best to stay engaged and interact with the streamer and other audience" (P5). Outside of the live streaming, they are very active in Wechat fan groups.

## RQ1: Why do female audiences watch these live streams by young male streamers on Douyin?

When asked what they like about their favorite streamer, most of them answered that the streamer has a nice voice (P9, P11, P13). This is unsurprising because all the participants are following streamers who show cast their voices while hiding their faces. One mentioned that it is because "he can sing" (P11), and one mentioned that, when listening to the streamer, she feels that "[her] whole person is quiet, and [her] temper has improved a lot" (P14). Only one mentioned that she "[doesn't] have too much feeling about his voice," she stays because the streamers and she are both eating lunch at that time (P5). Other reasons are also related to the streamer's characteristics, such as humorous (P13). Another important reason is that the streamer "does not use strategies [to ask for gifts]" (P11) or "demands for gifts" (P13).

We then asked the participants what made them keep coming back to the live stream room almost every day for a long while. One pointed out that the content of a live stream is unique; "if [the live stream] is missed today, it will be gone" (P5). Another often mentioned reason is long-term company. One told us that she "does not feel bored if there is a live stream to watch every day so that [she] can go and chat with other audiences in the room" (P11). This agrees with Lu et al. (2018), that they found "communication with others" is one of the important factors people engage with live streams.

We realize that these audiences watch live streams every day not just for the streamers, but also for the interaction and companionship they can gain. As P13 put, "[they] are happy when we gather together. [they] provide each other with [their] companions. The host is more like a bridge, but not the absolute protagonist" (P13). P11 elaborated that watching live streams has become a habit: "[i]t feels like you will slowly form a habit of being willing to go to that place, and after a long time, it will give you a sense of belonging. I think this is why there are so many people in each live stream room. The long-term company, that is because the mutual aura or magnetic field is more compatible with each other" (P11).

In summary, female audiences are attracted to a streamer because of his skills or characteristics, but they also value the interactions with other audiences. They seek companionship when watching live streams. Eventually, watching live streams becomes a habit.

## RQ2: Why do female audiences send virtual gifts to streamers?

Watching live streams is free, but since these Chinese streamers are making a living out of it, sending virtual gifts has become an unstated rule and a common practice. Although streamer as a profession is very common and there are various platforms that support live streaming, such as YouTube and Twitch, earning money through live streaming is a distinctive feature on Chinese live streaming apps (Zhang et al. 2019). A survey on live stream audiences on other platforms revealed that 66% of the audience have sent virtual gifts (Lu et al. 2018).

Therefore, it is not surprising that the most common answer we found was that they "send virtual gifts as a reward" (P11). Some may send a virtual gift when the streamer "sings a song that [she] like[s]" (P7, P9) or "touches [her]" (P11). P14 told us that she may send gifts when "[she] think[s] that he is worthwhile" (P14).

Streamers and Douyin are conscious of maximizing their virtual gift incomes. One participant told us that she may send virtual gifts when the streamer suggests the audiences do so (P9). Owing to the success of Douyin's UI design, some participants told us that sometimes they send a virtual gift "because [she] want[s] to unlock a certain special animation" (P7). Lu et al. (2018) also found that wanting to see fancy animations of virtual gifts is a common reason.

Audiences may choose not to send any gifts when watching a live stream. For example, P9 mentioned that she does not send gifts to popular streamers because "they don't need [her] gift at all" (P9). The behavior of watching live streams without sending any virtual gifts is called *Baipiao* (白嫖). The literal translation is "to visit a prostitute for free." Although this term is ridiculing the audience, it also implies that the streamers are "prostitutes," insinuating that they are offering services to satisfy audiences' carnal desire.

There are also voluntary cases. For example, P5 considers herself as a "girlfriend fan" because she sends gifts to make the streamer happy. She said that it is something she "might do to her boyfriend" (P5). P9 mentioned that if "[she hasn't] listened to the live stream for a long time, [she] will send it when [she] come[s]" (P9). These echo the long-term company they have found in live stream rooms.

The amount of virtual gifts they send has nothing to do with the quality of the content. All participants noted that it depends on their financial situation. Some send "more at the beginning of the month, less at the end of the month" (P14). It may also depend on "the degree to which [she] feel[s] sorry for him" (P5). It also depends on their other activities. For example, as P9 elaborated, if "[she] gave a gift of 100 Chinese Yuan today, it does not affect [her] mood for going out to eat hot pot and milk tea at night with [her] friends, nor does it affect [her] mood for shopping" (P9).

We also asked if they expect any reactions or privilege from the streamer. Most of them say they have no expectation (P5, P9, P13). Even if they hope for some appreciation, "it is not required" (P14), because "he will see it sooner or later and that's enough" (P9). No one expects any privileges when sending virtual gifts apart from showing appreciation, such as by putting their IDs on the screen (P7) or singing a song for them (P9, P14). Some of them are added to the streamer's personal Wechat account, but they do not consider it a privilege and it is not their purpose (P05).

In summary, sending virtual gifts as an appreciation is a common practice. Most of the time, the audience sends virtual gifts as a reward for the streamer's content. While the audience can choose not to send gifts, some of them are willing to send virtual gifts out of affection. They do not expect any returns from the streamer other than some verbal appreciation. This seemingly nonreciprocal exchange can be viewed as an appreciation of the streamer's companionship.

## RQ3: What's the relationship between the female audiences and the male streamers?

Many participants consider themselves as just a fan of the steamer. They consider their relationship the same as "the one between sales and consumers" (P14), so they "do not expect to be a true friend" (P9). These audiences do not anticipate connections outside of livestream rooms (P5). Although they are all in a Wechat fan group, P5 pointed out that she interacts more with the other audience members (P5). This echoes the point that the audience may find long-term companionship from other audiences.

Some participants consider themselves as Internet friends with the streamer. Some consider the steamer "the same as real-life friends" (P11) whereas some say there are still some differences because "the distance is too far" (P9). P11 pointed out that this depends on whether you can get along with the streamer, "if you can get along, then you can become friends. If you can't get along, then you just feel relaxed and happy in his live stream room" (P11).

In fact, some streamers and their audiences have developed very close relationships, albeit remotely. In some of the Wechat fan groups, the streamer may disclose their phone numbers and home addresses. One streamer shared his phone number with his dedicated fans and asked them to give him a morning call if he oversleeps. It is common among these fans to send physical gifts, such as food, clothes, paper towels, a computer, or a nice microphone. Lu et al. (2018) also found that 51% of their survey respondents had rewarded streamers by sending physical gifts or money via Wechat or AliPay.

In summary, we found that the relationship between the audience and the streamers can be either as consumers or as friends. Because the streamers the participants follow have a relatively smaller group of audience each day (fewer or around 20 active audience members), their live stream content is usually chatting, which helps improve their intimacy. In the long run, the audience may develop more personal emotions towards the streamers.

## Discussion

One major theme that emerges from the interviews is long-term companionship. Although many participants mentioned that they like the characteristics and skills of the streamer, more often they express their sense of belonging, which drives them to visit the live stream and send virtual gifts regularly. This is especially prominent last year. Due to the COVID19, many people had to stay at home. As a result, many of them started watching live streams to rid off loneliness.

Many participants consider themselves as consumers and streamers as sales or service providers. As a result, many of them consider sending virtual gifts as a reward or payment for their services or performances. However, we notice that the amount they send does not correlate with the quality of the live stream. For example, one may send virtual gifts because she cares for him or because she wants to compensate for the days she missed. It suggests that they are paying based on the emotional support they can gain from the streamers.

Related to long-term companionship, many participants value the friendship they have developed with not only the streamers but also other audiences. From the 6 interviews we have analyzed, it is not yet clear whether participants are using live stream rooms as an escape from or a substitute for reality. However, we noticed that P13 considers it as another social circle,
*"Watching live streams allows me to join another social circle, where there's no entanglement of profit from reality. People come here simply because they all like this streamer, have the same hobby. We are happy when we gather together. We provide each other with our companions. The host is more like a bridge, but not the absolute protagonist"* (P13).

Another theme is loyalty, which did not fit with any of the RQs above. Many participants follow only one streamer very closely and have a very high-level fan badge. For example, P5, branding herself as a girlfriend fan, only watches that particular streamer. The majority of the money P11 spent for her level 40 wealth badge goes to one streamer. These two fans are also very dedicated live stream room managers. Whenever the streamers start live streaming, they always join in right away. Throughout the entire 3+ hours, through typing, they welcome new audience members, show appreciation for gifts, kick out those unfriendly users, etc. When asked about why she has been willing to spend so much time and energy, she replies,
*" First of all, you must first agree with this live stream room. You like the atmosphere of the host, a kind of environment for getting along with some of the live stream rooms and other fans, that is, it feels very comfortable. People are willing to stay in a comfortable place, and there is another one, that is, over time, you will slowly develop a habit, because of this comfort. It feels like it will make you slowly form a habit of being willing to go to that place, and after a long time, it will give you a sense of belonging. I think this is why there are so many people in each live stream room. The*

*long-term company, that is because the mutual aura or magnetic field is more compatible with each other*" (P11).

## Limitations

There are many limitations to this study. One is that we have limited ways of recruiting participants. As a result, the participants and the streamers whom our participants follow are not very diverse. So far, the results can not explain some of the situations the author witnessed in some live stream rooms. For example, one female audience criticizes that it is too easy to become top 1 on the streamer's leaderboard and demands the other audiences to send more gifts to make it more challenging. However, this may be a rare case. A more common case is that the audience often is curious about the streamer's height, weight, and looks if the streamer does not show his face while streaming. Why is the streamer's appearance important? Do fans of streamers who show their faces have different preferences?

Another limitation is that interview data is not sufficient in understanding these audience's behavior. Observational data from live stream rooms and contents in Wechat fan groups may also be helpful.

## Future directions

The current future plan includes conducting more interviews with diverse audiences who watch a diverse range of streamers. More analysis can be done from the perspectives of the female consumers, gender relationships, and socioeconomics.

Reference

Corbin, J., & Strauss, A. (2014). *Basics of qualitative research: Techniques and procedures for developing grounded theory.* Sage publications.

Kaye, D. B. V., Chen, X., & Zeng, J. (2020). The co-evolution of two Chinese mobile short video apps: Parallel platformization of Douyin and TikTok. *Mobile Media & Communication*, 2050157920952120.

Li, B., Guan, Z., Chong, A. Y. L., & Hou, F. (2018). What drives people to purchase virtual gifts in live streaming? The mediating role of flow.

Li, Xiaomeng. (2020). "How Powerful Is the Female Gaze? The Implication of Using Male Celebrities for Promoting Female Cosmetics in China." *Global Media and China* 5 (1): 55–68. https://doi.org/10.1177/2059436419899166.

Lu, X., & Lu, Z. (2019, July). Fifteen seconds of fame: A qualitative study of Douyin, a short video sharing mobile application in China. In *International Conference on human-computer interaction* (pp. 233-244). Springer, Cham.

Lu, Z., Xia, H., Heo, S., & Wigdor, D. (2018, April). You watch, you give, and you engage: a study of live streaming practices in China. In *Proceedings of the 2018 CHI conference on human factors in computing systems* (pp. 1-13).

Qu, T. (2020, April 30). TikTok and China version Douyin surpass 2 billion download milestone, underlining continued appeal. *South China Morning Post*. https://www.scmp.com/tech/apps-social/article/3082285/tiktok-and-china-version-douyin-surpass-2-billion-download

Xiaoxing Zhang, Yu Xiang & Lei Hao (2019): Virtual gifting on China's live streaming platforms: hijacking the online gift economy, Chinese Journal of Communication, DOI: 10.1080/17544750.2019.1583260

13